\documentclass[showpacs,preprintnumbers,final,twocolumn]{revtex4}%
\usepackage{amssymb}
\usepackage{amsmath}
\usepackage{graphicx}
\usepackage{dcolumn}
\usepackage{bm}
\usepackage{amsfonts}
\usepackage{setspace}%
\providecommand{\U}[1]{\protect\rule{.1in}{.1in}}
\begin{document}
\title{Hysteresis and intermittency in a nano-bridge based suspended DC-SQUID}
\author{Eran Segev}
\author{Oren Suchoi}
\author{Oleg Shtempluck}
\author{Fei Xue}
\altaffiliation{Current affiliation: Department of Physics, University of Basel.}

\author{Eyal Buks}
\affiliation{Department of electrical engineering, Technion, Haifa 32000, Israel}
\date{\today }

\begin{abstract}
We study voltage response of nano-bridge based DC-SQUID fabricated on a
$\operatorname{Si}_{\mathrm{3}}\mathrm{N}_{\mathrm{4}}$ membrane. Such a
configuration may help in reducing $1/f$ noise, which originates from
substrate fluctuating defects. We find that the poor thermal coupling between
the DC-SQUID and the substrate leads to strong hysteretic response of the
SQUID, even though it is biased by an alternating current. In addition, when
the DC-SQUID is biased near a threshold of spontaneous oscillations, the
measured voltage has an intermittent pattern, which depends on the applied
magnetic flux through the SQUID.

\end{abstract}
\maketitle

Superconducting quantum interference devices (SQUIDs) are key components in
many applications \cite{Clarke2004}. One of the major research goals with
SQUIDs nowadays is to reduce noises in the Josephson junctions (JJs) composing
the SQUIDs. This issue was extensively studied in the past decades, where it
was shown that one of the main sources of noise is the coupling of the JJs to
the substrate, which induces $1/f$ noises \cite{Zorin1996_13682}. Several
studies have tried to solve this problem by suspending the JJs and thus
decouple them from the substrate. So far these studies have only shown
moderate success \cite{Krupenin1998_3212,Krupenin2000_287,Li2007_033107}.
Another emerging technology, which attracts an increasing interest, is SQUIDs
based on nano-bridge weak-links. Such SQUIDs can be made extremely small and
have the potential to outperform conventional SQUIDs in terms of noise
properties \cite{Vijay2010_223112a,Vijay2009_087003}.

Our original research goal was to study noise properties of a nano-bridge
based DC-SQUIDs\ suspended on a Silicon Nitride ($\operatorname{Si}%
_{\mathrm{3}}\mathrm{N}_{\mathrm{4}}$) membrane. However, we discovered that
non-equilibrium thermal processes, which emerge due to the poor coupling of
the JJs to the substrate, play a dominant role in SQUIDs dynamics. When the
SQUID is excited by alternating current strong hysteretic response of the
DC-SQUID is observed, even though such excitation should, in principle,
eliminate hysteretic behavior. In addition, when the SQUID is biased near a
threshold of spontaneous oscillations, the measured voltage shows an
intermittent pattern, which depends on the magnetic flux through the SQUID. To
account for these results we extend the theoretical model in Ref.
\cite{Segev2010_arxiv} to include low frequency variations in the SQUID
temperature. Numerical simulations show qualitative agreement with the
experimental data.%

\begin{figure}
[ptb]
\begin{center}
\includegraphics[
height=3.0245in,
width=3.1789in
]%
{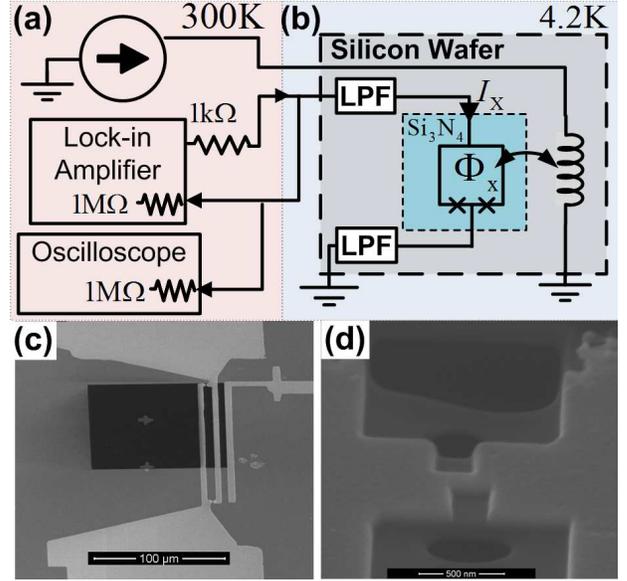}%
\caption{(Color online) $\mathrm{(a)}$ Measurement setup. $\mathrm{(b)}$ A
simplified circuit layout for a DC-SQUID. $\mathrm{(c)}$ Electron micrograph
of a DC-SQUID partially fabricated on top a $\operatorname{Si}_{\mathrm{3}%
}\mathrm{N}_{\mathrm{4}}$ membrane. $\mathrm{(c)}$ Electron micrograph image
of a nano-bridge.}%
\label{ExpSetup}%
\end{center}
\end{figure}

A simplified circuit layout of a typical device is illustrated in Fig.
\ref{ExpSetup}$(\mathrm{b})$. We build our devices on a high resistivity
Silicon wafer, covered by a $100%
\operatorname{nm}%
$-thick layer of $\operatorname{Si}_{\mathrm{3}}\mathrm{N}_{\mathrm{4}}$. A
small section of the wafer is etched from the back to produce a suspended
$\operatorname{Si}_{\mathrm{3}}\mathrm{N}_{\mathrm{4}}$ membrane, having size
of $100%
\operatorname{\mu m}%
^{2}$. The DC-SQUID is made of\ Niobium, having layer thickness of $80%
\operatorname{nm}%
$ and lateral dimensions of $110\times7%
\operatorname{\mu m}%
^{2}$.\ The self-inductance of the SQUID is calculated using FastHenry program
\cite{Kamon94fasthenry} to be $L=100\mathrm{pH}$. The DC-SQUID is composed of
two nano-bridge JJs (NBJJs), one NBJJ in each of its two arms. The dimensions
of the bridges are $100\times115%
\operatorname{nm}%
^{2}$, and their combined critical current is $1.2%
\operatorname{mA}%
$. Most of the SQUID structure, including one of the NBJJs is fabricated on
the $\operatorname{Si}_{\mathrm{3}}\mathrm{N}_{\mathrm{4}}$ membrane, as shown
by the electron micrographs in subfigures \ref{ExpSetup}$\mathrm{(c)}$ and
\ref{ExpSetup}$\mathrm{(d)}$. An on-chip stripline passes nearby the DC-SQUID,
and is used to generate magnetic flux in the DC-SQUID loop. A DC bias line,
which includes on-chip low-pass filters (LPFs), is connected to the DC-SQUID
and is used for voltage measurements of the SQUID. Note that the DC-SQUID is
embedded in a superconducting stripline resonator (not illustrated in Fig.
\ref{ExpSetup}$\mathrm{(b)}$), but the resonator was not used in the
experiments, and thus had a negligible influence on the experiments presented
in this paper. Further design considerations and fabrication details can be
found elsewhere \cite{Segev2009_152509,Suchoi2010_174525}.

Our experiments are carried out using the setup depicted in Fig.
\ref{ExpSetup}$(\mathrm{a})$. We use a lock-in amplifier, which applies
alternating current through the SQUID. The excitation frequencies can be up to
$100%
\operatorname{kHz}%
$. We measure the voltage across the DC-SQUID using the lock-in amplifier and
record its time domain dynamics using an oscilloscope. In experiments showing
intermittency we also apply DC magnetic flux through the SQUID. All
measurements are carried out while the device is fully immersed in liquid Helium.%

\begin{figure}
[ptb]
\begin{center}
\includegraphics[
height=2.642in,
width=3.3719in
]%
{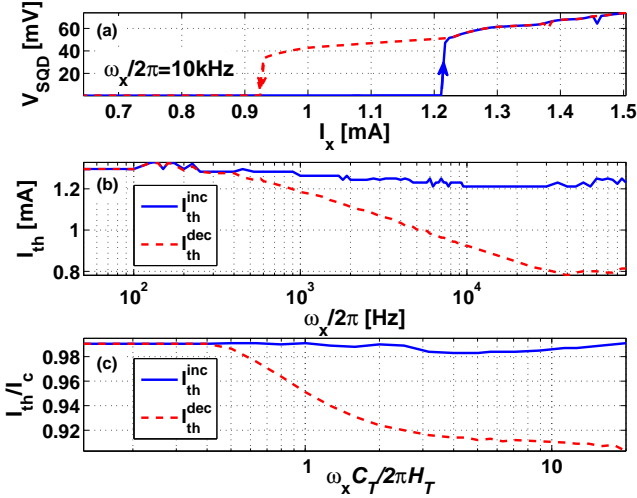}%
\caption{Hysteretic current-voltage measurements. Panel $\mathrm{(a)}$ shows
DC-SQUID\ voltage traces measured for increasing and decreasing current
excitation sweeps. Panels $\mathrm{(b)}$ and $\mathrm{(c)}$ show measurement
and simulation results, respectively, of increasing and decreasing threshold
currents as a function of the excitation frequency.}%
\label{HystIV}%
\end{center}
\end{figure}

Figure \ref{HystIV}$\mathrm{(a)}$ shows an example of the DC-SQUID hysteretic
response to an excitation current having alternating frequency of
$\omega_{\mathrm{x}}/2\pi=10%
\operatorname{kHz}%
$. This measurement shows the measured voltage across the SQUID when the
current amplitude is swept up and then down. Each recorded point is averaged
over $100%
\operatorname{ms}%
$, thus including $1000$ excitation cycles. Such alternating excitation
should, in principle, eliminate hysteresis in the response of the DC-SQUID,
provided that all characteristic time scales of the device are much shorter
than the excitation period. Note that, the thermal relaxation rate of a local
hot-spot, created in a nano-bridge having good thermal coupling to the
substrate, is typically on the order of gigahertz \cite{Tarkhov2008_241112}.
In contrary to our expectation, the measurement of hysteretic current-voltage
characteristic behavior under alternating current excitation indicates that,
there is a much longer characteristic thermal time-scale, on the order of
millisecond, which affects the DC-SQUID dynamics. Such a long time-scale may
exist in our device because the $\mathrm{Si}_{\mathrm{3}}\mathrm{N}%
_{\mathrm{4}}$ membrane, to which the SQUID is coupled, has relatively large
heat capacity, and also because of the poor coupling between this membrane and
its surrounding, which results in a relatively long thermal relaxation time.

Figure \ref{HystIV}$\mathrm{(b)}$ summarizes several measurements like the one
shown in Fig. \ref{HystIV}$\mathrm{(a)}$. It plots the increasing and
decreasing threshold currents (i.e. the values of applied currents
corresponding to voltage jumps in the increasing and decreasing current sweeps
respectively) as a function of the lock-in amplifier carrier frequency. These
threshold currents are approximately equal one another only up to frequencies
of about $500%
\operatorname{Hz}%
$. The hysteretic nature of the DC-SQUID emerges when the excitation frequency
further increases. The decreasing threshold current falls to lower values
while the increasing one only slightly degraded. This trend continues up to
frequencies of about $40%
\operatorname{kHz}%
$, where both threshold currents cease to depend on the excitation frequency.

In order to account for the above results, we extend the model in Ref.
\cite{Segev2010_arxiv}\ of hysteretic, metastable DC-SQUID to include slow
variations of the SQUID temperature. According to the model, the equations of
motion DC-SQUID are the following two equations for the NBJJs\ phases
$\gamma_{1}$ and $\gamma_{2}$
\begin{gather}
\ddot{\gamma}_{1}+\beta_{\mathrm{D}}\dot{\gamma}_{1}+\left(  1+\alpha
_{0}\right)  y\left(  \Theta\right)  \sin\gamma_{1}\nonumber\\
+\frac{1}{\beta_{\mathrm{L0}}}\left(  \gamma_{1}-\gamma_{2}+2\pi
\Phi_{\mathrm{x}}/\Phi_{0}\right)  =I_{\mathrm{x}}/I_{\mathrm{c}%
0}+g_{\mathrm{n}},\label{eom gamma_1}%
\end{gather}%
\begin{gather}
\ddot{\gamma}_{2}+\beta_{\mathrm{D}}\dot{\gamma}_{2}+\left(  1-\alpha
_{0}\right)  y\left(  \Theta\right)  \sin\gamma_{2}\nonumber\\
-\frac{1}{\beta_{\mathrm{L0}}}\left(  \gamma_{1}-\gamma_{2}+2\pi
\Phi_{\mathrm{x}}/\Phi_{0}\right)  =I_{\mathrm{x}}/I_{\mathrm{c}%
0}+g_{\mathrm{n}},\label{eom gamma_2}%
\end{gather}
where the overdot denotes a derivative with respect to a normalized time
parameter $\tau=\omega_{\mathrm{pl}}t$, where $\omega_{\mathrm{pl}}$ is the
DC-SQUID plasma frequency. In addition, $I_{\mathrm{x}}$ is the bias current,
$\Phi_{\mathrm{x}}$ is the external magnetic flux, $\Phi_{0}$ is flux quantum,
$\beta_{\mathrm{D}}$ is the dimensionless damping coefficient, and $E_{0}$ is
the Josephson energy. The critical current of the DC-SQUID is temperature
dependant, thus we employ the notation $I_{\mathrm{c0}}$ for the
SQUID\ critical current at the base temperature $T_{0},$ defined by the
coolant. The dimensionless parameters $\beta_{\mathrm{L0}}$ and $\alpha_{0}$
characterize the DC-SQUID hysteresis and asymmetry at the base temperature
$T_{0}$, respectively. The dimensionless factor $g_{\mathrm{n}}$ is a noise
term, which is neglected in our numerical calculations.

Assuming that, the temperature of the SQUID, $T$, has no spatial dependence.
The term $y\left(  \Theta\right)  $ expresses the dependence of the
NBJJs\ critical currents on the temperature. It is given by $y\left(
\Theta\right)  \equiv\widetilde{y}\left(  \Theta\right)  /\widetilde{y}\left(
\Theta_{0}\right)  $ \cite{Skocpol1976_1045}, where $\Theta=T/T_{\mathrm{c}}$,
$\Theta_{0}=T_{0}/T_{\mathrm{c}}\,$, and $T_{\mathrm{c}}$ is the critical
temperature of the DC-SQUID. The function $\widetilde{y}$ is given by
$\widetilde{y}\left(  \Theta\right)  =\left(  1-\Theta^{2}\right)
^{3/2}\left(  1+\Theta^{2}\right)  ^{1/2}$. Using the notation $\beta
_{\mathrm{C}}=2\pi C_{\mathrm{T}}T_{\mathrm{c}}/\Phi_{0}I_{\mathrm{c}0}$ and
$\beta_{\mathrm{H}}=H_{\mathrm{T}}/C_{\mathrm{T}}\omega_{\mathrm{pl}}$, where
$C_{\mathrm{T}}$ is thermal heat capacity and $H_{\mathrm{T}}$ is heat
transfer rate, the SQUID heat balance equation reads \ %

\begin{equation}
\dot{\Theta}=\frac{\beta_{\mathrm{D}}}{\beta_{\mathrm{C}}}\left(  \dot{\gamma
}_{1}^{2}+\dot{\gamma}_{2}^{2}\right)  -\beta_{\mathrm{H}}\left(
\Theta-\Theta_{0}\right)  \;, \label{EOM_Theta}%
\end{equation}

Simulation results showing the dependence of the hysteretic behavior of the
DC-SQUID on the excitation frequency are shown in Fig. \ref{HystIV}%
$\mathrm{(c)}$. This panel shows the increasing and decreasing threshold
currents as a function of the excitation frequency, normalized by the thermal
relaxation time. These results show qualitative agreement with the
corresponding experimental results shown in Fig. \ref{HystIV}$\mathrm{(b)}$.%

\begin{figure}
[ptb]
\begin{center}
\includegraphics[
height=2.7181in,
width=3.5111in
]%
{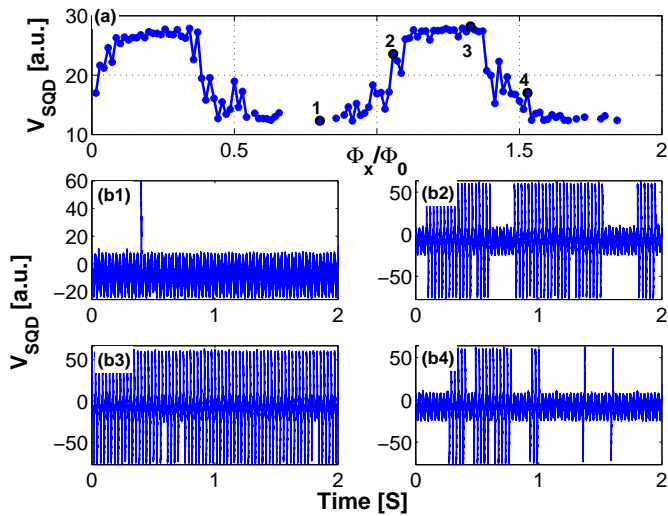}%
\caption{Intermittent behavior. $\mathrm{(a)}$ Measured average squared
voltage of the DC-SQUID. $\mathrm{(b}i{\protect\normalsize )}$ Time traces of
the DC-SQUID voltage taken at the working points marked by the corresponding
number $i$ in panel $\mathrm{(a)}$.}%
\label{Intermirrency}%
\end{center}
\end{figure}

Figure \ref{Intermirrency} shows intermittent behavior of the SQUID. In this
measurement the DC-SQUID is biased near the threshold of spontaneous
oscillations using alternating current having frequency of $\omega
_{\mathrm{x}}/2\pi=70%
\operatorname{Hz}%
$. The voltage across the SQUID is measured as a function of time for various
values of magnetic flux applies through the SQUID. Panel $\mathrm{(a)}%
$\textrm{ }shows the average value of the squared voltage versus the magnetic
flux. Panels $\mathrm{(b}i\mathrm{)}$ show voltage time traces measured for
the corresponding marked points in panel $\mathrm{(a)}$. The low amplitude
voltage oscillations are measured due to the existence of a parasitic serial
resistance in the measurement wiring. Each time the alternating current drives
the SQUID into the oscillatory zone it responses with a voltage spike measured
on top of the low amplitude parasitic voltage. Spikes can be positive or
negative depending on the polarity of the driving current. The measured traces
show intermittent behavior, in which the spikes are generated in bunches. This
behavior suggests that in this region the system becomes thermally bistable.
The bistability is realized by switching between the cold locally stable
state, where the DC-SQUID is in the superconducting state, and the hot locally
stable state, where the SQUID becomes normal. Switching between these states
is randomly triggered by external noise.

In conclusion, we have studied nano-bridge based DC-SQUIDs fabricated on a
$\operatorname{Si}_{\mathrm{3}}\mathrm{N}_{\mathrm{4}}$ membrane. We find that
the response of the DC-SQUID to a low frequency alternating excitation current
is hysteretic, and that the strength of the hysteresis depends on the
excitation frequency. In addition, when the SQUID is biased near a threshold
of spontaneous oscillations, the measured voltage has an intermittent pattern,
which depends on the applied magnetic flux through the SQUID. Such hysteretic
response degrades the performance of the suspended SQUID, and questions the
effectiveness of using suspension as a method for noise reduction.

E.S. is supported by the Adams Fellowship Program of the Israel Academy of
Sciences and Humanities. This work is supported by the German Israel
Foundation under grant 1-2038.1114.07, the Israel Science Foundation under
grant 1380021, the Deborah Foundation, Russell Berrie nanotechnology
institute, Israeli Ministry of Science, the European STREP QNEMS project, and MAFAT.

\bibliographystyle{apsrev}
\bibliography{Bibilography}

\end{document}